 \documentclass[traditabstract]{aa}
\usepackage{natbib,twoopt,amssymb,lscape}
\usepackage[breaklinks=true]{hyperref} 
\bibpunct{(}{)}{;}{a}{}{,} 
\newcommandtwoopt{\citeads}[3][][]{\href{http://adsabs.harvard.edu/abs/#3}%
{\citealp[#1][#2]{#3}}} 
\newcommandtwoopt{\citepads}[3][][]{\href{http://adsabs.harvard.edu/abs/#3}%
{\citep[#1][#2]{#3}}} 
\newcommandtwoopt{\citetads}[3][][]{\href{http://adsabs.harvard.edu/abs/#3}%
{\citet[#1][#2]{#3}}} 
\newcommandtwoopt{\citeyearads}[3][][]%
{\href{http://adsabs.harvard.edu/abs/#3}{\citeyear[#1][#2]{#3}}}
\usepackage{graphicx}
\usepackage{longtable}
\usepackage{natbib}
\bibpunct{(}{)}{;}{a}{}{,} 
\usepackage{txfonts}

\makeatletter

\newcommand{\Rmnum}[1]{\expandafter\@slowromancap\romannumeral #1@}

\newcommand{\MSUN}{M$_\odot$\space}
\newcommand{\MSUNnospace}{M$_\odot$}
\makeatother
\begin{document}

\title{Testing the cooling flow model in the intermediate polar EX~Hydrae}

   \author{G. J. M. Luna,
          \inst{1,2}
          J. C. Raymond,
          \inst{2}
          N. S. Brickhouse,
          \inst{2}
          C.~W. Mauche,
          \inst{3}
          \and
          V. Suleimanov
          \inst{4,5}
          }

   \institute{Instituto de Astronom\'ia y F\'isica del Espacio (IAFE, CONICET-UBA), CC 67 - Suc. 28 (C1428ZAA)  CABA -- Argentina.\\
              \email{gjmluna@iafe.uba.ar}
\and
 Harvard-Smithsonian Center for Astrophysics, 60 Garden st., Cambridge, MA, 02138
\and
Lawrence Livermore National Laboratory, L-473, 7000 East Avenue, Livermore, CA 94550, USA
\and
Institute for Astronomy and Astrophysics, Kepler Center for Astro and Particle Physics, Eberhard Karls University, Sand 1, D-72076 T\"ubingen, Germany
\and
Kazan (Volga Region) Federal University, Kremlevskaya 18, 420008 Kazan, Russia
}
   \date{}

 \abstract
{We use the best available X-ray data from the intermediate polar EX~Hydrae to study the cooling-flow model often applied to interpret the X-ray spectra of these accreting magnetic white dwarf binaries.  First, we resolve a long-standing discrepancy between the X-ray and optical determinations of the mass of the white dwarf in EX~Hya by applying new models of the inner disk truncation radius. Our fits to the X-ray spectrum now agree with the white dwarf mass of 0.79 \MSUN determined using dynamical methods through spectroscopic observations of the secondary. We use a simple isobaric cooling flow model to derive the emission line fluxes, emission measure distribution, and H-like to He-like line ratios for comparison with the 496 ks {\it Chandra\/} High Energy Transmission Grating observation of EX~Hydrae. We find that the H/He ratios are not well reproduced by this simple isobaric cooling flow model and show that while H-like line fluxes can be accurately predicted, fluxes of lower-Z He-like lines are significantly underestimated.  This discrepancy suggests that an extra heating mechanism plays an important role at the base of the accretion column, where cooler ions form.  We thus explored more complex cooling models, including the change of gravitational potential with height in the accretion column and a magnetic dipole geometry. None of these modifications to the standard cooling flow model are able to reproduce the observed line ratios.  While a cooling flow model with 
subsolar (0.1 $\odot$) abundances is able to reproduce the line ratios by reducing the cooling rate at temperatures lower than $\sim 10^{7.3}$ K, the predicted line-to-continuum ratios are much lower than observed.  We discuss and discard mechanisms, such as photoionization, departures from constant pressure, resonant scattering, different electron-ion temperatures, and Compton cooling.  Thermal conduction transfers energy from the region above $10^7$ K, where the H-like lines are mostly formed, to the cooler regions where the He-like ions of the lower-Z elements are formed, hence in principle it could help  resolve the problem.  However, simple models indicate that the energy is deposited below $10^6$ K, which is too cool to increase the emission of the He-like lines we observe.  We conclude that some other effect, such as thermally unstable cooling, modifies the temperature distribution.}


\titlerunning{Testing the cooling flow model.}
   \maketitle
%

\section{Introduction}
\label{sec:intro}

Cooling flows have been postulated to occur in a diversity of settings, from clusters of galaxies, where the low-density, intracluster medium loses its gravitational energy by radiating X-rays \citepads{1977MNRAS.180..479F,1981ApJ...247..464B}, to accreting magnetic and nonmagnetic white dwarfs (WDs) in binary systems. In nonmagnetic WDs, such as dwarf novae \citepads{1985ApJ...292..535P} and symbiotics \citepads{2007ApJ...671..741L}, a cooling flow is present in the boundary layer of the accretion disk, where material has to dissipate its rotational energy before settling onto the WD surface. If the WD magnetic field is strong enough to disrupt the accretion disk, the material flowing through the magnetically-dominated accretion column reaches highly supersonic velocities and undergoes a strong shock near the WD surface, heating the plasma to keV temperatures; the material in the postshock region cools until it settles onto the WD surface.

In intermediate polars (IPs), the WD has a  magnetic field  strong enough to disrupt the accretion disk at the Alfv\'{e}n radius $R_{\rm in}$ \citepads{2011MNRAS.411.1317R,2014arXiv1405.0855S}, where magnetic pressure balances the ram pressure of the accretion flow. The WD's magnetic field then channels the material from the inner disk toward the magnetic poles. Most of the cooling radiation from the accretion column is emitted at X-ray wavelengths, predominantly as bremsstrahlung and line emission, i.e., a cooling flow.

For galaxy clusters, spatially resolved X-ray imaging observations are available, and surface brightness and temperature profiles have been compared with the predictions of the cooling flow model. The observations do not show the large amounts of cool X-ray emitting plasma at the center of the clusters predicted by the cooling flow model. This discrepancy is known as the {\it \textup{cooling flow problem}\/} \citepads{2006PhR...427....1P}. As the X-ray emitting region of the accreting WD cannot be spatially resolved, an analogous cooling flow problem has been more difficult to discern. Hints of problems with the simple cooling flow model came from residuals in the fits of low- and medium-resolution X-ray spectra. However, these residuals can be reduced by invoking complex absorption \citepads[e.g.,]{2008MNRAS.387.1157R,2010A&A...520A..25Y}. Further progress has been stymied by the low resolution of CCD spectra and the limited number and quality of high-resolution grating spectra. \citetads{2003ApJ...586L..77M} modeled the high-resolution {\it Chandra\/} X-ray spectra of EX Hydrae (EX~Hya), two dwarf novae, and an old nova using a solar abundance cooling flow model, matching the line fluxes to within a factor of two. On the other hand, fits with the cooling flow model to the high-resolution X-ray spectra of the dwarf nova WX~Hyi \citepads{2003ApJ...598..545P} and the likely magnetic system V426~Oph \citepads{2004ApJ...610..991H} showed good agreement with the high temperature 
part of 
the spectrum, but underpredicted emission from the cooler part of the flow. This discrepancy appears to be in  opposite direction from that seen in galaxy clusters.  It is possible that there is additional heat as the gas cools to $10^7$ K in both cases, but that in EX Hya this heating simply increases the
emission measure of the million degree gas, whereas in clusters the heating is strong enough to prevent the gas from cooling to  lower temperatures.

Depending on the physical properties of the binary system, additional heating and cooling mechanisms can play important roles in the energy equation of the cooling flow; e.g., gravity is a source of extra thermal energy when the shock height is a significant fraction of the WD radius. The cooling mechanism has been the subject of many theoretical studies \citepads[e.g.,][]{1973PThPh..49.1184A,1999MNRAS.306..684C,2005A&A...440..185C,2014MNRAS.438.2267H}, and the derived spectral models have shown reasonable success in tests on low- and medium-resolution X-ray spectra. Given that the shock temperature is proportional to the strength of the gravitational potential well, its measurement can be used to determine the mass of the accreting WD. \citetads{1991ApJ...367..270I} modeled the X-ray spectra of IPs observed with {\it GINGA\/} as bremsstrahlung emission and used the derived temperatures to determine the masses of the accreting WDs.  \citetads{1999MNRAS.306..684C} also derived WD masses from {\it GINGA\/} data, but considered additional terms in the energy equation of the cooling flow, such as gravity, as well as modifications of the observed spectra by reflection by the WD surface and absorption due to a warm absorber above the emission region. \citetads{2005A&A...443..291S} and \citetads{2009A&A...496..121B} modeled {\it RXTE\/} and {\it Swift\/}/BAT spectra of a sample of IPs using the \citetads{1999MNRAS.306..684C} model.  \citetads{2014MNRAS.438.2267H} combined the 
magnetic field geometry, varying gravitational potential, nonequipartition between ions and electrons, and nonequilibrium ionization to derive a model for the postshock region of the accretion column, which was recently tested on {\it Suzaku\/} data of EX~Hya \citepads{2014MNRAS.441.3718H}.

Given that the cooling plasma also produces line emission, \citetads{1997ApJ...474..774F} realized that the then-higher spectral resolution of {\it ASCA\/} CCD spectra could be used to determine the shock temperature, and hence the WD mass, through measurements of the line intensity ratios of the H-like to He-like ions.  They assumed the temperature and density structures given by \citetads{1973PThPh..49.1184A}, where cooling is due only to bremsstrahlung radiative losses.  Although line fluxes and equivalent widths from Fe, Ar, S, Si, and Mg were measured, Mg lines were not used to determine the WD mass because they are blended with Fe L-shell lines at the {\it ASCA\/}/SIS spectral resolution.

Are the more recent, high-quality spectra obtained with the {\it Chandra\/} High-Energy Transmission Grating (HETG) compatible with the models proposed for the cooling of the accretion column in EX~Hya? \citetads{2010ApJ...711.1333L} discovered high-velocity wings on several bright emission lines in the 496 ks HETG spectrum of EX~Hya, which they attributed to photoionized gas in the accretion column above the shock. Since only a small fraction of the X-ray spectrum originates in the photoionized preshock material, with the rest attributed to collisionally ionized plasma, a cooling flow model of the postshock emission should give a good first-order match to the observed spectrum.

In Section \ref{sec:exhya} we describe the properties of EX~Hya. We solve the long-standing controversy about the WD mass of EX~Hya derived using X-ray spectral fits compared with optical/IR radial velocity measurements. We use the 496 ks {\it Chandra\/} HETG spectrum described in Section \ref{sec:obs} to test models of the cooling in the postshock region (Section \ref{sec:model}). We first determine the shock temperature, which matches the value implied by a WD mass of 0.79 \MSUNnospace, from the least-absorbed spin-phase in Section \ref{sec:tshock}. We then utilize the H-like to He-like line ratios as a proxy for the temperature and density structure of the column. We start by computing a simple isobaric cooling flow spectrum due to bremsstrahlung and line emission. We compute the emission measure distribution (EMD), line fluxes, and line ratios (Section \ref{sec:cf}).  We find that simple isobaric cooling flow models are not able to reproduce the observed spectrum and line ratios. Specifically, we find that the H-like line fluxes are well reproduced, whereas the same model underestimates the He-like line fluxes. This implies that either heat is not efficiently removed in the upper portion of the accretion column, or there is an extra source of heat in the lower portion of the accretion column.  We then consider the effects on the EMD of a different geometrical configuration by calculating a magnetic dipole model (Section \ref{sec:dipole}). We find that the line ratios 
predictions are mostly unaffected by the accretion column geometry. In Section \ref{sec:abund} we compute subsolar abundance models. While models with very low abundances are able to reproduce the observed line ratios, they greatly underestimate the observed line-to-continuum ratios.

Motivated by recent studies of the importance of thermal conduction in the cooling of the intracluster medium of clusters of galaxies \citepads[e.g.,][]{2003ApJ...596..889K,2011MNRAS.414.1493R} and WD boundary layers \citepads{2008A&A...483..231L}, we consider thermal conduction in the energy equation of the flow in the accretion column of EX~Hya (Section \ref{sec:conduction}).  While this changes the EMD in the correct sense, the change is too small to alleviate the discrepancy.  We conclude that another effect, such as thermally unstable cooling, must be responsible for the excess emission in the $10^6$ to $10^7$ K range.  In Section \ref{sec:disc} we discuss our findings and present concluding remarks. In the Appendix we discuss  various approximations in the simple cooling model.

\section{EX Hya}
\label{sec:exhya}

EX~Hya is an IP, a magnetic cataclysmic variable (CV) whose magnetic field is strong enough to channel the accretion flow onto the magnetic poles, but not strong enough to entirely disrupt the accretion disk.  The orbital and WD spin periods are 98.26 and 67.03 minutes, respectively.  EX Hya has been extensively studied at optical, UV, EUV, and X-ray wavelengths \citepads[e.g.,][]{1988MNRAS.231..549R,1999ApJ...520..822M,2003ApJ...586L..77M, 2009A&A...496..121B}. The luminosity and accretion rate are estimated to be $(2.6\pm0.6) \times 10^{32}~\rm erg~s^{-1}$ and $(6.2\pm1.5) \times 10^{-11}~\rm (M_{\rm WD}/0.5~M_{\odot})^{-1.61}~M_{\odot}$ yr$^{-1}$ at a distance of 64.5 pc \citepads{2003A&A...412..821B}.  There has been a long-standing discrepancy between the WD mass of $0.790\pm 0.026$ \MSUN derived from the radial velocities of optical and IR emission lines \citepads{2008A&A...480..199B} and masses of 0.42--0.66 \MSUN derived by equating the X-ray temperature to the shock temperature expected for free fall onto the WD \citepads[e.g.,][]{2010A&A...520A..25Y}.  

Various models have been used to fit the X-ray spectrum of EX~Hya, and different values for the WD masses have been obtained. \citetads{2010A&A...520A..25Y} modeled the {\it Suzaku\/} X-ray spectrum including gravity in the flow equations and obtained a mass of 0.42$\pm$0.02 \MSUNnospace; \citetads{2005A&A...443..291S} modeled the {\it RXTE\/} X-ray spectrum also including gravity in the flow equations and obtained a mass of 0.50$\pm$0.05 \MSUNnospace; \citetads{2009A&A...496..121B} obtained a mass of 0.66$\pm$0.17 \MSUN fitting the {\it Swift\/}/BAT spectrum and using the same model as \citetads{2005A&A...443..291S}. \citetads{2014MNRAS.441.3718H} combined the magnetic field geometry, varying gravitational potential, nonequipartition between ions and electrons, and ionization nonequilibrium to develop a model that was applied to {\it Suzaku\/} data of EX~Hya, and obtained a mass of  0.63$^{+0.17}_{-0.14}$ \MSUNnospace. \citetads{1997ApJ...474..774F} measured the ratio of the intensities of H-like to He-like emission lines in their study of the {\it ASCA\/} spectrum of EX~Hya and obtained a mass of 0.48$_{-0.06}^{+0.10}$ \MSUNnospace. Radial velocity measurements of the X-ray emission lines observed in a short (50 ks) {\it Chandra\/} observation, combined with velocity amplitude $K_2$ of the secondary derived by \citetads{2003MNRAS.342..151V} led \citetads{2004ApJ...610..411H} to derive a mass of 0.49$\pm$0.13 \MSUNnospace. Dynamical methods through optical and near-IR 
spectroscopic observations of the secondary by \citetads{2008A&A...480..199B}, significantly improved the measurements of $K_2$ and concluded that the mass of the WD is 0.790$\pm$0.026 \MSUNnospace. The WD mass derived by \citetads{2004ApJ...610..411H} matches the value derived by \citetads{2008A&A...480..199B} using their more accurate measurement of $K_{2}$.

All the previous determinations of the WD mass based on the shock temperature of EX~Hya, assumed that the material in the accretion column was at free-fall speed from infinity.  However, in systems such as EX~Hya, the inner radius of the accretion disk can be close to the WD surface \citepads{2011MNRAS.411.1317R} and therefore a factor that takes this into account has to be introduced in the free-fall speed equation as

\begin{equation}
v_{\rm ff}=\sqrt{2 G M_{\rm WD}  \left(\frac{1}{R_{\rm WD}+h_{\rm sh}}- \frac{1}{R_{\rm in}}\right)}\ ,
\label{eq:ff}
\end{equation}

\noindent
where $R_{\rm WD}$ is the radius of the WD and 
$h_{\rm sh}$ is the height of the accretion column. If we use $M_{\rm WD}$=0.79 $M_{\odot}$, $R_{\rm WD}$=7$\times$10$^{8}$ cm, $h_{\rm sh}$=0.1$\times R_{\rm WD}$, and $R_{\rm in}$=2.7$\times R_{\rm WD}$ \citepads{2011MNRAS.411.1317R,2014arXiv1405.0855S}, we have $v_{\rm ff} =$ 4000 km s$^{-1}$. Using the strong shock condition $kT_{\rm shock}=3\mu m_{\rm p} v_{\rm ff}^{2}/16$, we find that $kT_{\rm shock} \sim 19$ keV.  \citetads{2003ApJ...586L..77M} obtained a good fit to a short {\it Chandra\/} HETG observation with a shock temperature of 20 keV, \citetads{2010A&A...520A..25Y} obtained $kT_{\rm shock}$=12.7$_{-0.9}^{+0.9}$ keV from their fit to {\it Suzaku\/} data, \citetads{2009A&A...496..121B} fit of {\it Swift\/}/BAT data yielded $kT_{\rm shock}$=19.4$_{-4.4}^{+4.4}$ keV, whereas \citetads{1997ApJ...474..774F} obtained $kT_{\rm shock}$=15.4$^{+5.3}_{-2.6}$ keV from the line ratios in the {\it ASCA\/} spectrum.

We therefore use a mass of 0.79 M$_\odot$ and a radius of $7 \times 10^8$ cm for the WD. The preshock density is above $10^{15}~\rm cm^{-3}$ \citepads{2010ApJ...711.1333L}, which, combined with the low magnetic field strength ($\sim 4\times 10^4$~G), means that cyclotron cooling can be neglected.

\section{Observations}
\label{sec:obs}

\begin{figure*}
\begin{center}
\includegraphics[scale=0.95]{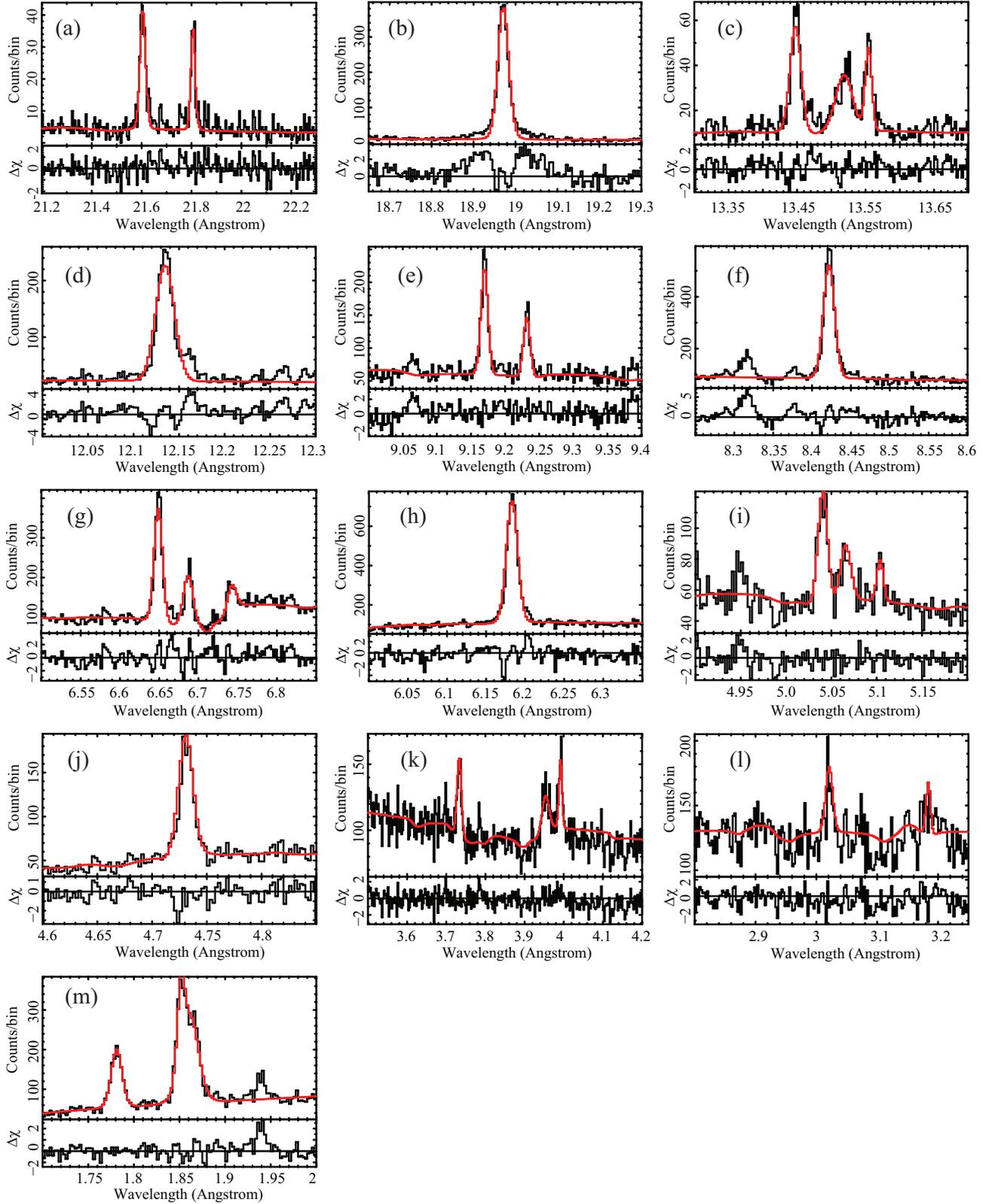}
\caption{Observed emission-line profiles with overlaid Gaussian models of
(a) O VII    $\lambda$21.6   and O VII $\lambda$21.8  (MEG $\pm$1);
(b) O VIII   $\lambda$18.97  (HEG $\pm$1);
(c) Ne IX    $\lambda$13.45  and Ne IX $\lambda$13.55 (HEG $\pm$1); 
(d) Ne X     $\lambda$12.13  (HEG $\pm$1); 
(e) Mg XI    $\lambda$9.16   and Mg XI $\lambda$9.23  (HEG $\pm$1);
(f) Mg XII   $\lambda$8.42   (HEG $\pm$1);
(g) Si XIII  $\lambda$6.64,   Si XIII  $\lambda$6.68, and Si XIII $\lambda$6.74  (HEG $\pm$1); 
(h) Si XIV   $\lambda$6.18   (HEG$\pm$1);
(i) S XV     $\lambda$5.04,   S XV     $\lambda$5.06, and S XV    $\lambda$5.10  (HEG $\pm$1); 
(j) S XVI    $\lambda$4.73   (HEG $\pm$1); 
(k) Ar XVIII $\lambda$3.73,   Ar XVII  $\lambda$3.944 + Ar XVII  $\lambda$3.961 + Ar XVII $\lambda$3.965, and 
                              Ar XVII  $\lambda$3.994 + S XVI    $\lambda$3.992 + S XVI   $\lambda$3.991 (HEG $\pm$1); 
(l) Ca XX    $\lambda$3.02  and Ca XIX $\lambda$3.17  (HEG $\pm$1);  and
(m) Fe XXVI  $\lambda$1.78,  Fe XXV    $\lambda$1.85  and Fe XXV $\lambda$1.86  (HEG $\pm$1),
where the HETG
data and model profiles are shown with black and red histograms, respectively. Below each panel, we plot the fit residuals. Where overlapping, the HEG$\pm$1 and MEG$\pm$1 spectra were fit simultaneously; HEG$\pm$1 is shown for illustrative purposes. As discussed in \citetads{2010ApJ...711.1333L}, the broad wings in the residuals to the fits of some of the emission lines --- O VIII in particular --- are due to emission by the photoionized preshock accretion flow.}
\label{fig:gauss}
\end{center}
\end{figure*}

EX~Hya was observed with {\it Chandra\/} using the ACIS-S/HETG combination for 496 ks. The observation was obtained in four segments (ObsIDs 7449: start time 2007 May 13 22:15:35 UT, exposure time 130.65 ks; 7452: start time 2007 May 17 03:12:38 UT, exposure time 49.17 ks, 7450: start time 2007 May 18 21:56:57 UT, exposure time 162.73 ks; and 7451: start time 2007 May 21 14:15:08 UT, exposure time 153.07 ks). We reduced the data using the {\it Chandra\/} Interactive Analysis of Observations (CIAO v.4.1) software package. Beyond the reduction of the {\it Chandra\/} data set already described in \citetads{2010ApJ...711.1333L}, we combined the +1 and $-$1 orders of the HEG and MEG arms using the script \texttt{add\_grating\_orders}. The HEG$\pm$1 and MEG$\pm$1 spectra were thus fit simultaneously. Line fluxes were measured by using Gaussians to represent the emission lines and a first-order polynomial to represent the nearby continuum (see Figure \ref{fig:gauss}).  We estimate that the measured fluxes differ by less than 2\%  using the more complex $\beta$-profile, which indeed was loaded with the RMF during the fit.

Table \ref{tab:2} lists the measured fluxes of individual H-like and He-like lines.  For the analysis discussed below, H-like Ly$\alpha$ and the He-like resonance ($r$), intercombination ($i$), and forbidden ($f$) lines are required. For O, Ne, and Mg, the $f$ lines are not observed. The $r$ and $i$ lines of Ar cannot be resolved by the HEG and these two lines are measured together using a single Gaussian component. Because the emissivity of the $f$ line of Ar XVII $\lambda$3.994 peaks at $\log{T}=$7.3, while the emissivities of S XVI $\lambda$3.991 and $\lambda$3.992 peak at $\log{T}=$7.4 (AtomDB), the measured flux of the line at $\lambda$3.99 consists of the summed contribution of these three lines, and the models take this into account. Since the He-like $r$, $i$, and $f$ lines of Ca XIX cannot be resolved, we used a single Gaussian  to measure these lines. Finally, the $i$ and $f$ lines of Fe XXIV cannot be resolved and thus we measured these two lines together using a single Gaussian. As discussed by \citetads{2010ApJ...711.1333L}, the broad wings in the residuals to the fits of some of the emission lines, in particular O VIII, are due to emission by the photoionized preshock accretion flow. Dielectronic recombination transitions, as well as other satellite lines located close enough to the He and H-like lines that cannot be resolved at the HEG spectral resolution were included in the measured fluxes and in the models described below. We found that they contribute to a small fraction ( a few percent) of the line fluxes.

\section{The postshock structure}
\label{sec:model}

\subsection{Determining the shock temperature}
\label{sec:tshock}

To avoid the complications of modeling the complex absorber in an X-ray global fit, we exploited the high quality of the data set and filtered the data to obtain a spectrum at spin-phase maximum when the observer has a virtually unblocked view of the accretion flow.  We divided the data into five spin-phase bins of width
$\Delta\phi_{67}$=0.2 ($\phi_{67}$ refers to the phases at the spin period
of 67.03 min) and fit the cooling flow model available in XSPEC (\texttt{mkcflow})
to the five spectra. We had already filtered in the binary
phase range $\phi_{98}$=0.0--0.6 ($\phi_{98}$ refers to the phases at the orbital period of 98.26 min) to avoid the absorption by the accretion disk bulge \citepads{2006ApJ...643L..45H}.   As expected from observations at EUV wavelengths \citepads{1997ApJ...477..390H,1999ApJ...520..822M} and from our spin-phase filter, the result of fitting the data in the spin-phase range $\phi_{67}$=0.9--1.1 yields a value for the absorption column $N_{\rm H}$ smaller than 10$^{20}$ cm$^{-2}$ and $\dot{M}=2.18_{-0.01}^{+0.01}\times 10^{-11}$ $M_{\odot}$ yr$^{-1}$. This fit yields a value of $kT_{\rm shock}$=25.5$_{-0.5}^{+0.6}$ keV. If we allow the elemental abundances to vary (using the XSPEC model \texttt{vmcflow}), we get a temperature $kT_{\rm shock}$=19.7$_{-0.1}^{+0.2}$ keV, $\dot{M}=2.55_{-0.01}^{+0.01}\times 10^{-11}$ $M_{\odot}$ yr$^{-1}$, and abundances \citepads[in units of the solar abundances of][]{1989GeCoA..53..197A} of O=0.92$_{-0.07}^{+0.05}$, Ne=1.31$_{-0.06}^{+0.06}$, Ar=0.93$_{-0.25}^{+0.22}$, Ca=1.07$_{-0.37}^{+0.31}$, Fe=0.61$_{-0.01}^{+0.01}$; Mg, S, and Si are consistent with solar values \citepads[these abundances agree with the values derived by][]{1997ApJ...474..774F}. Although these abundances look less anomalous in units of other modern solar 
abundance determinations \citepads[e.g.,][]{2009LanB...4B...44L}, we retain the \citetads{1989GeCoA..53..197A} scale because AtomDB and XSPEC use them as defaults. The temperature derived is thus that predicted using eq. \ref{eq:ff} and the strong shock condition with a WD mass of 0.79 M$_{\odot}$, in agreement with values determined dynamically. Therefore, throughout the following sections, we  adopt a shock temperature $kT$=19.7 keV and the abundances derived from the \texttt{vmcflow} model.

\begin{figure*}
\begin{center}
\includegraphics[scale=0.6]{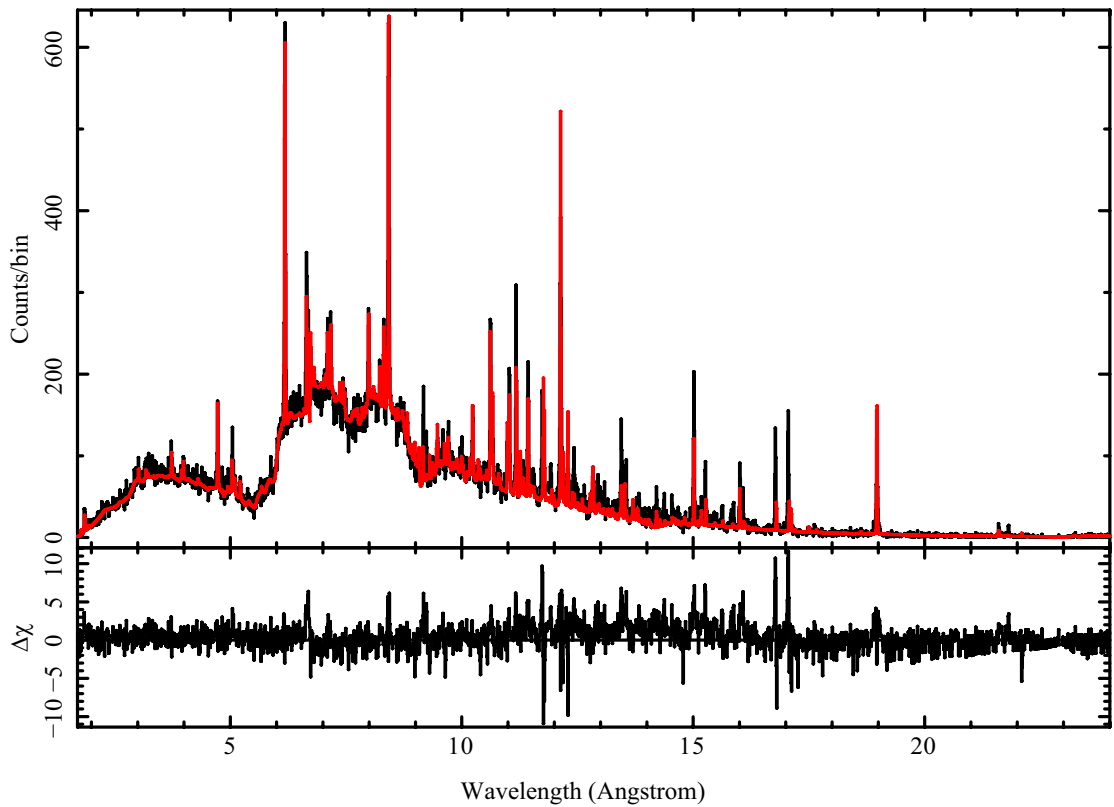}
\caption{{\it Chandra\/} MEG $\pm$1 order X-ray spectrum between spin-phases $\phi_{67}$=0.9 and 1.1 (i.e., the range of phases where the observer has the most unextinguished view of the accretion column) fit with an XSPEC variable-abundance isobaric cooling flow model (\texttt{vmcflow}). The absorption column, $N_{\rm H}$, is smaller than 10$^{20}$ cm$^{-2}$, as expected, while the shock temperature is $kT_{\rm shock}$=19.7$_{-0.1}^{+0.2}$ keV.}
\label{fig:spinmax}
\end{center}
\end{figure*}

\subsection{The isobaric cooling flow. Derivation of a physically-based emission measure distribution}
\label{sec:cf}

\begin{figure}
\includegraphics[scale=0.5]{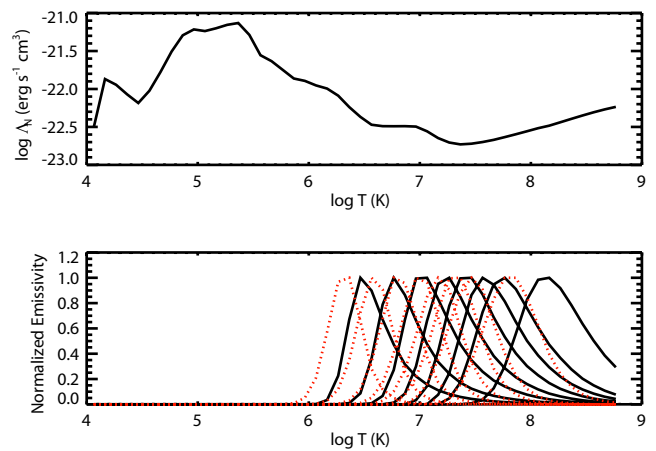}
\caption{{\it Upper panel\/}: Radiative cooling function derived using emissivities from AtomDB (in the low-density limit, see Section \ref{sec:cf}) and abundances \citepads[in units of the solar abundance of][]{1989GeCoA..53..197A} of O=0.92, Ne=1.31, Ar=0.93, Ca=1.07, Fe=0.61 (see Section \ref{sec:tshock}). {\it Lower panel\/}: Line emissivities as a function of temperature from AtomDB, normalized by their peak emissivity, for the H-like (full, black line) and He-like resonance+intercombination (dotted, red line) transitions of O, Ne, Mg, Si, S, Ar, Ca, and Fe.}
\label{fig:cf}
\end{figure}

In the simplest scenario, the plasma in the accretion column radiates its excess energy in an isobaric flow where the cooling is determined by the radiative cooling function $\Lambda (T)$ (Fig. \ref{fig:cf}). The cooling time can be expressed as

\begin{equation}
 t_{\rm cool}=\frac{5kT}{n\Lambda (T)}\ ,
\label{eq13}
\end{equation}

\noindent
where the density $n=n_{0}T_{0}/T$, with the $0$ indicating values just beneath the shock. The rate of temperature change can be expressed as

\begin{equation}
\frac{dT}{dt}=\frac{n_{0}T_{0}\Lambda(T)}{5k_{\rm B}T}\ ,
\label{eq28}
\end{equation}

\noindent
where $k_{\rm B}$ is the Boltzmann's constant.

The speed in the cooling flow $v=dx/dt$ can be written as

\begin{equation}
v=\frac{v_{0} n_0}{n}
\label{eq17}
\end{equation}

\noindent
from mass flux conservation, where $v_{0}=v_{\rm ff}/4$. Then, at each place in the column, we can write

\begin{equation}
 dx=v dt=\frac{v_{\rm ff}n_{0}}{4 n}\frac{5k_{\rm B}T}{n_{0}T_{0}\Lambda(T)}dT=\frac{5k_{\rm B}T^{2}v_{\rm ff}}{4 T_{0}^{2}n_{0}\Lambda(T)}dT \ .
\label{eq30}
\end{equation}

\noindent
The emission measure (EM) is defined as $EM=\int{n^2 dV}$. A simple geometrical model for the accretion column is a cylinder, with a radius $r_{\rm spot}$ and height $h_{\rm sh}$.  In a cylinder, the area element $dV/dx=\pi r^2$ is constant, and using eq. \ref{eq30} and $n=n_{0}T_{0}/T$, we have the following expression for the EM: 
\begin{equation}
EM=\frac{5}{4}\pi r_{spot}^{2} n_{0}v_{\rm ff}k_{\rm B}\int_{\Delta T}{\Lambda(T)^{-1} dT}\ ,
\label{eq34} 
\end{equation}

\noindent
where $\Delta T$ is the temperature range from the shock temperature to the minimum temperature ($kT_{min}$) of the gas  when it reaches the WD surface. In practice, the value of $kT_{min}$ is set by the lowest temperature for which the cooling function was calculated, which in our case is $\log{T}$=4. In fact, the assumption that the gas can be described in the coronal approximation, where the ions are in the ground state, breaks down below $\log{T}$=6, but that region makes no significant contribution to the X-ray emission lines discussed here (see Fig. \ref{fig:cf}).

The flux of a given emission line is 

\begin{equation}
f_{\lambda}=\frac{A}{4\pi d^2}\int_{\Delta T}{\epsilon(T)\times EMD(T) dT}\ ,
\label{eq31}
\end{equation}

\noindent
where $A$ is the elemental abundance, a constant that modifies the solar abundances \citepads{1989GeCoA..53..197A} already embedded in $\epsilon(T)$, the emissivity of the ion from AtomDB \citepads{2012ApJ...756..128F}, EMD(T) is the distribution of the emission measure as function of temperature (i.e., the $Emission Measure Distribution$); note that $\Lambda(T)$ was computed with the set of elemental abundances found in the global spectral fit in Section \ref{sec:tshock}. 

While most of the constants in equation~\ref{eq34} are unknown, they primarily affect the absolute intensities of the emission lines, so the ratios of H-like to He-like emission lines are more easily compared with observations \citepads{1997ApJ...474..774F}. The constants in equation~\ref{eq34}  affect the EMD, and hence the line ratios indirectly, in that they determine the shock height and therefore  contributions of the change in gravitational potential and  compression due to dipole geometry of the postshock flow. In the present case, we neglect the dipole geometry and assume a shock height much less than the WD radius. From equation~\ref{eq31}, we can express the H to He line ratio as

\begin{equation}
\frac{\rm H}{\rm He}=\frac{\int{\epsilon_{Ly\alpha}(T)\times \Lambda(T)^{-1} dT}} {\int{(\epsilon_{He_{r}}+\epsilon_{He_{i}}+\epsilon_{He_{f}})\times \Lambda(T)^{-1} dT}}\,,
\label{eq27}
\end{equation}

\noindent
where the subindices $r$, $i$, and $f$ refer to the resonance, intercombination, and forbidden transitions, respectively, and Ly$\alpha$ includes contributions from both components of the doublet. Using equations \ref{eq34} and \ref{eq27} with $kT_{\rm shock}$=19.7 keV, we calculated the H/He line ratios for O, Ne, Mg, Si, S, Ar, Ca, and Fe and compared them with the ratios measured in the {\it Chandra\/} data (see Table \ref{tab:1}). The observed ratios of the intensities of Fe XVII and Fe XXII lines \citepads{2001ApJ...560..992M,2003ApJ...588L.101M} indicate that the density in the accretion column is very high, with $n \gtrsim$ 10$^{14}$ cm$^{-3}$. At these densities, the sum of the flux of the $i$ and $f$ lines is nearly constant, with the $f$ line flux suppressed in favor of the $i$ line flux as the density increases. Thus, the lack of observed $f$ lines for O, Ne, and Mg should have a negligible effect on the sum of the observed He-like line fluxes. The $r$ line is very weakly density dependent. Using the sum also reduces density-dependent effects.  We use the public AtomDB emissivities, computed for $n_{\rm e}$=1 cm$^{-3}$, and thus although the flux of the forbidden lines at this low density is much higher than the flux of the intercombination lines, their sum is density independent. 

\begin{figure*}
\begin{center}
\includegraphics[scale=0.8]{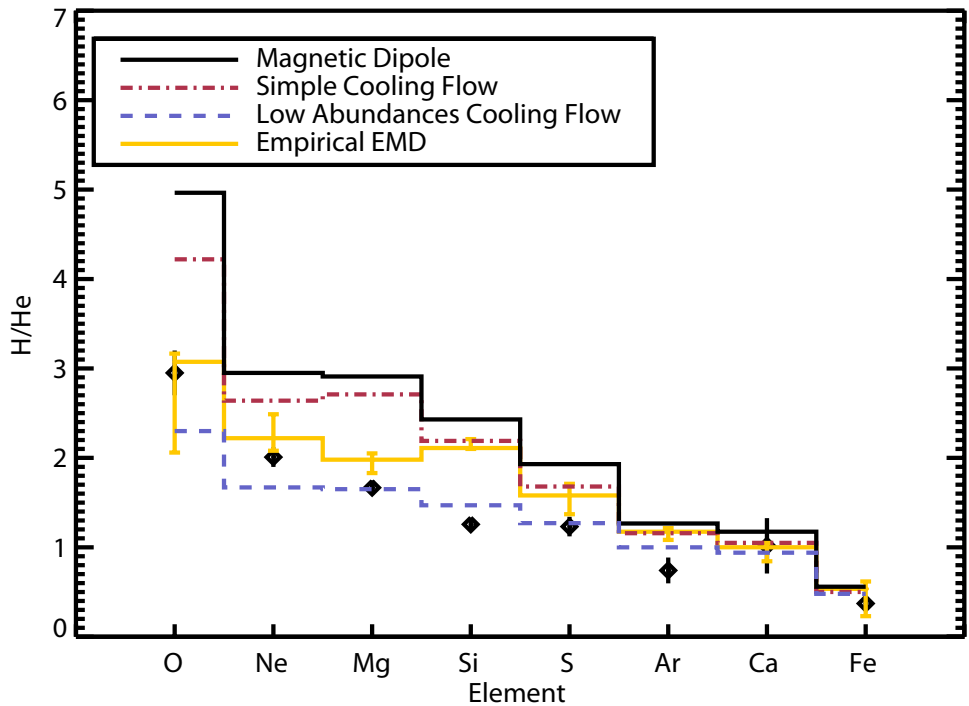}
\caption{Measured ratios of the H-like Ly$\alpha$ line to the sum of the He-like resonance, intercombination, and forbidden lines from the {\it Chandra\/} spectrum of EX Hya ({\it black diamonds with error bars\/}). The histograms show the results from the models discussed in Sections \ref{sec:cf} to \ref{sec:conduction}. For the empirically-derived EMD, the error bars indicate the 90\% confidence level (Sect. \ref{sec:poa})}
\label{fig:ratios}
\end{center}
\end{figure*}

In Fig. \ref{fig:ratios} we plot the measured line ratios and model predictions. The isobaric cooling flow model overestimates the line ratios for ions with Z $<$ 14. As the emissivities of H- and He-like ions peak at different temperatures (Fig. \ref{fig:cf}), they are emitted at different heights in the accretion column, and thus the observed disagreement in the line ratios suggests that the temperature and density structures of the accretion column are not adequately described by this simple model. Although the measured ratios were not corrected for absorption, the difference in the absorption cross section between the H- and He-like lines of the same ion is so small that low-Z lines, such as O~VII $\lambda$21.60 and O~VIII $\lambda$18.97, would have to be attenuated by orders of magnitude to change the line ratios by factors of 2.

To determine which physical mechanism could be missing in our simple model, we need to determine whether the model  fails to reproduce  the H-like line fluxes, the He-like line fluxes, or both. We can use equation \ref{eq31} to calculate the line fluxes if we know the values of the constants in equation \ref{eq34}: the radius of the accretion spot $r_{\rm spot}$ and the free-fall speed $v_{\rm ff}$, which in turn is determined by the shock temperature.  We assume $v_{\rm ff}$ as derived in Section \ref{sec:tshock} and $r_{\rm spot}=\sqrt{\dot{M}/\pi v_{\rm ff} n_{0} m_{\rm H}}$. The best match between data and model is obtained assuming an accretion rate 
$\dot{M}$=1.74$\times$10$^{-11}$~\rm $M_{\odot}$ yr$^{-1}$ (note that Beuermann et al.\ 2003 estimated $\dot{M}$=2.94$\times$10$^{-11}$~\rm $M_{\odot}$ yr$^{-1}$ for $M_{\rm WD}=0.79$ \MSUNnospace). The cooling flow model closely reproduces the H-like lines fluxes, whereas it underestimates the He-like lines fluxes (Table \ref{tab:2}). 

Using the specific accretion rate of $a$=3 g cm$^{-2}$ s$^{-1}$ and $f$=1.6$\times$10$^{-4}$, as recently obtained by \citetads{2014arXiv1405.0855S}, resulted in both H- and He-like fluxes overestimated by more than an order of magnitude.  \citetads{1997ApJ...474..774F} found that their cooling flow model fit the line ratios of high-Z ions (Z $>$ 14), assuming that the bottom of the accretion column has a temperature $kT_{\rm B}$=0.65 keV, preventing any cooling below this temperature. In this scenario, the H-like line intensities are well reproduced but the He-like lines of low-Z ions (Mg, Ne, O) are significantly underpredicted. 

\renewcommand{\arraystretch}{1.3}
\begin{table*}[h]
\caption[]{H- and He-like line fluxes (in 10$^{-4}$ photons s$^{-1}$ cm$^{-2}$) measured and derived from the models discussed in Section \ref{sec:model}.} 
\centering
\begin{tabular}{l c c c c c}
 \hline\hline
Ion  & Data & Model A\tablefootmark{a} & Model B\tablefootmark{b}& Model C\tablefootmark{c}& Model D\tablefootmark{d}\\
\hline
\hbox to 2.5in{O VII    $\lambda$21.8   \leaders\hbox to 0.4em{\hss.\hss}\hfill}& 1.21$_{-0.14}^{+0.23}$& 0.92&  0.85&  0.45 &   0.40--1.53\\
\hbox to 2.5in{O VII    $\lambda$21.6   \leaders\hbox to 0.4em{\hss.\hss}\hfill}& 1.66$_{-0.17}^{+0.27}$& 1.11&  1.03&  0.53 &   0.48--1.59\\
\hbox to 2.5in{O VIII   $\lambda$18.97  \leaders\hbox to 0.4em{\hss.\hss}\hfill}& 8.49$_{-0.31}^{+0.31}$& 8.57&  9.35&  2.24 &   3.45--9.33\\
Ne IX    $\lambda$13.55 + Fe XIX $\lambda$13.551 + Fe XX $\lambda$13.535 + $\lambda$13.533  & 0.56$_{-0.07}^{+0.06}$& 0.54&  0.53&  0.14 &  0.25--0.71\\
\hbox to 2.5in{Ne IX    $\lambda$13.45 + Fe XIX $\lambda$13.462 \leaders\hbox to 0.4em{\hss.\hss}\hfill}& 1.47$_{-0.08}^{+0.08}$&  0.73&   0.75&   0.20 &   0.36--1.00\\
\hbox to 2.5in{Ne X     $\lambda$12.13 + Fe XVII $\lambda$12.124 \leaders\hbox to 0.4em{\hss.\hss}\hfill}& 4.07$_{-0.17}^{+0.09}$&  3.36&  3.82&  0.58 &    1.49--4.01\\
\hbox to 2.5in{Mg XI    $\lambda$9.23 + Fe XXII $\lambda$9.231  \leaders\hbox to 0.4em{\hss.\hss}\hfill}& 0.34$_{-0.03}^{+0.02}$& 0.26&   0.28&  0.05 &   0.13--0.36\\
\hbox to 2.5in{Mg XI    $\lambda$9.16   \leaders\hbox to 0.4em{\hss.\hss}\hfill}& 0.57$_{-0.03}^{+0.03}$& 0.36&  0.39&  0.07 &  0.17--0.47\\
\hbox to 2.5in{Mg XII   $\lambda$8.42   \leaders\hbox to 0.4em{\hss.\hss}\hfill}& 1.51$_{-0.04}^{+0.03}$& 1.70&  1.97&  0.21 &   0.65--1.72\\
\hbox to 2.5in{Si XIII  $\lambda$6.74   \leaders\hbox to 0.4em{\hss.\hss}\hfill}& 0.20$_{-0.02}^{+0.02}$& $\cdots$&$\cdots$ & $\cdots$& $\cdots$ \\
\hbox to 2.5in{Si XIII  $\lambda$6.68 + $\lambda$6.663 + $\lambda$6.664  \leaders\hbox to 0.4em{\hss.\hss}\hfill}& 0.45$_{-0.04}^{+0.05}$& 0.39&  0.40&  0.06 &  0.16--0.40\\
\hbox to 2.5in{Si XIII  $\lambda$6.64   \leaders\hbox to 0.4em{\hss.\hss}\hfill}& 0.83$_{-0.03}^{+0.03}$& 0.59&  0.63&  0.09 &  0.24--0.59\\
\hbox to 2.5in{Si XIV   $\lambda$6.18   \leaders\hbox to 0.4em{\hss.\hss}\hfill}& 1.86$_{-0.04}^{+0.03}$& 2.14&  2.54&  0.23 &  0.83--2.22\\
\hbox to 2.5in{S XV     $\lambda$5.10 + $\lambda$5.098 + $\lambda$5.103  \leaders\hbox to 0.4em{\hss.\hss}\hfill}& 0.09$_{-0.02}^{+0.03}$& $\cdots$&$\cdots$ &$\cdots$ &$\cdots$ \\
\hbox to 2.5in{S XV     $\lambda$5.06 + $\lambda$5.050  \leaders\hbox to 0.4em{\hss.\hss}\hfill}& 0.27$_{-0.06}^{+0.06}$& 0.33&  0.34&  0.03 &  0.10--0.28\\
\hbox to 2.5in{S XV     $\lambda$5.04   \leaders\hbox to 0.4em{\hss.\hss}\hfill}& 0.48$_{-0.04}^{+0.05}$& 0.40&  0.44&  0.05 &   0.17--0.42\\
\hbox to 2.5in{S XVI    $\lambda$4.73   \leaders\hbox to 0.4em{\hss.\hss}\hfill}& 1.01$_{-0.05}^{+0.05}$& 1.14&  1.41&  0.11 &    0.43--1.19\\
\hbox to 2.5in{Ar XVII  $\lambda$3.944 + 
                        $\lambda$3.961 +
                        $\lambda$3.965  \leaders\hbox to 0.4em{\hss.\hss}\hfill}& 0.14$_{-0.04}^{+0.03}$& 0.11&  0.18& 0.01&     0.05--0.12\\ 
\hbox to 2.5in{Ar XVII  $\lambda$3.994 +
                S XVI   $\lambda$3.991 +
                        $\lambda$3.992  \leaders\hbox to 0.4em{\hss.\hss}\hfill}& 0.18$_{-0.03}^{+0.03}$& $\cdots$& $\cdots$&$\cdots$&$\cdots$ \\ 
\hbox to 2.5in{Ar XVIII $\lambda$3.73   \leaders\hbox to 0.4em{\hss.\hss}\hfill}& 0.23$_{-0.03}^{+0.04}$& 0.19&  0.25&  0.02 &   0.08--0.22\\ 
\hbox to 2.5in{Ca XIX   $\lambda$3.184 +
                        $\lambda$3.211  \leaders\hbox to 0.4em{\hss.\hss}\hfill}& 0.17$_{-0.03}^{+0.04}$& 0.13&  0.16&  0.02&    0.05--0.14\\
\hbox to 2.5in{Ca XX    $\lambda$3.02   \leaders\hbox to 0.4em{\hss.\hss}\hfill}& 0.18$_{-0.05}^{+0.03}$& 0.14&  0.19&  0.02&    0.04--0.14\\
\hbox to 2.5in{Fe XXV   $\lambda$1.85   \leaders\hbox to 0.4em{\hss.\hss}\hfill}& 1.96$_{-0.40}^{+0.29}$& 1.05&  1.37&  0.21 &   0.35--0.97\\
\hbox to 2.5in{Fe XXV   $\lambda$1.865 + 
                        $\lambda$1.868  \leaders\hbox to 0.4em{\hss.\hss}\hfill}& 1.47$_{-0.29}^{+0.42}$& 0.68&  0.88&  0.13 &   0.25--0.67\\
\hbox to 2.5in{Fe XXVI  $\lambda$1.78   \leaders\hbox to 0.4em{\hss.\hss}\hfill}& 1.27$_{-0.10}^{+0.11}$&  0.87&  1.26&  0.16 &   0.13--1.02\\ 
\hline
\label{tab:2}
\end{tabular}
\tablefoot{
\tablefoottext{a}{Cooling flow model. Geometry of the accretion column: cylinder. No gravity. Solar abundances. Fractional area $f$=10$^{-4}$, accretion rate $\dot{M}$=1.74$\times$10$^{-11}$~\rm $M_{\odot}$ yr$^{-1}$. }
\tablefoottext{b}{Suleimanov et al.\ (2005) model. Geometry of the accretion column: magnetic dipole. Solar abundances. Accretion rate $\dot{M}$=1.74$\times$10$^{-11}$~\rm $M_{\odot}$ yr$^{-1}$,
                  specific accretion rate $a$=0.6 g cm$^{-2}$ s$^{-1}$.}
\tablefoottext{c}{Cooling flow model. Geometry of the accretion column: cylinder. Abundances 0.1 solar. Fractional area $f$=10$^{-4}$, accretion rate $\dot{M}$=1.74$\times$10$^{-11}$~\rm $M_{\odot}$ yr$^{-1}$.}
\tablefoottext{d}{EMD derived empirically from the measured flux ratios of H- and He-like lines of O, Ne, Mg, Si, S, Ar, Ca, and Fe. Error bars were determined at the 90\% confidence level.}}

\end{table*}

\begin{figure*}
\begin{center}
\includegraphics[scale=0.8]{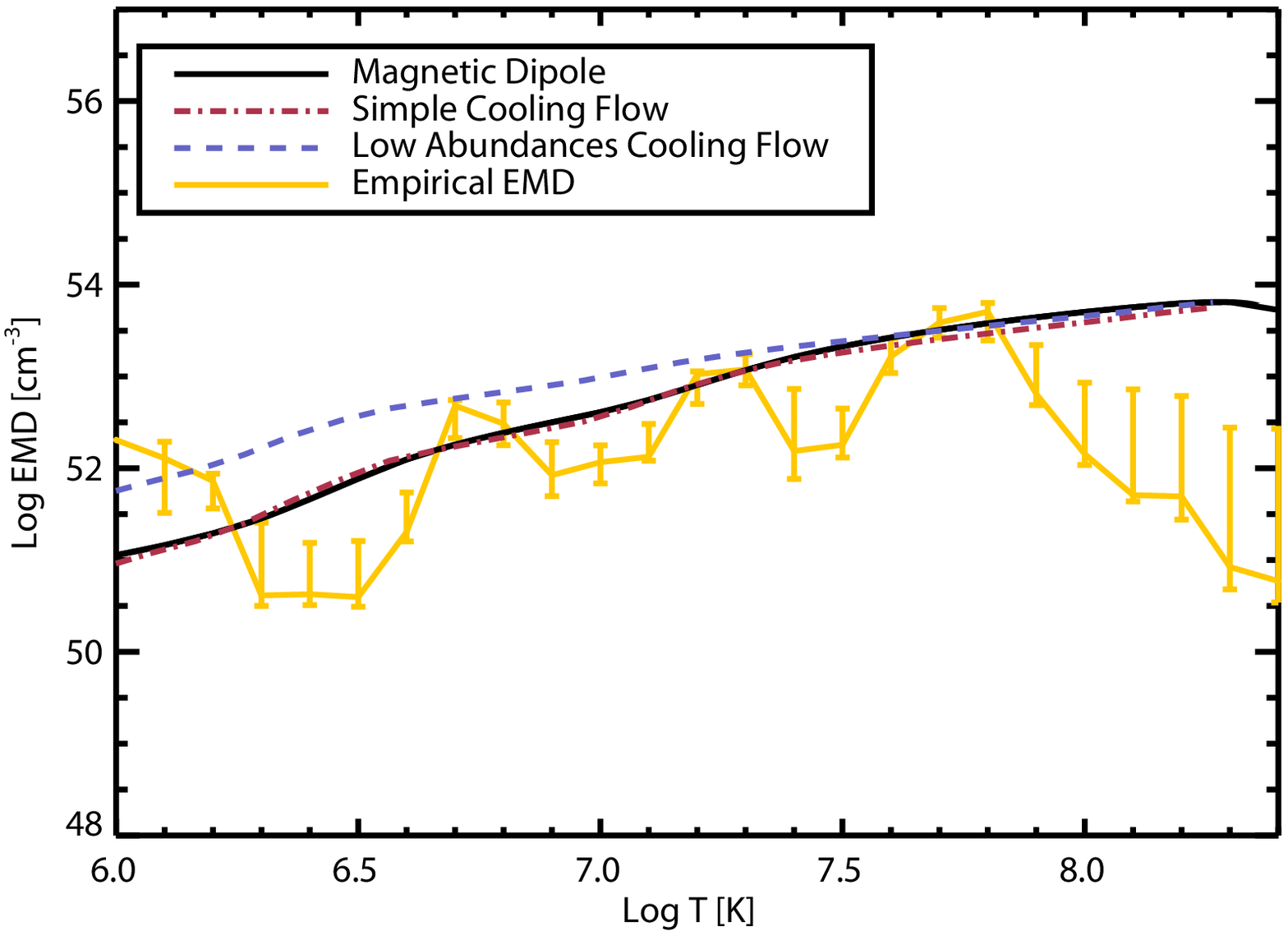}
\caption{Emission measure distributions (EMDs) derived from models of the cooling of the postshock plasma discussed in Sections \ref{sec:cf} to \ref{sec:conduction}. Error bars indicate the 90\% confidence level of the empirically-derived EMD (Sect. \ref{sec:poa}). }
\label{fig:emd}
\end{center}
\end{figure*}

\subsection{Magnetic dipole geometry} 
\label{sec:dipole}

The simple cooling flow model (Model A in Table \ref{tab:1}) makes a number of approximations, neglecting photoionization, departures from constant pressure, departures from planar geometry, variations of the gravitational potential with height, resonant scattering, thermal conduction, electron-ion temperature differences, and Compton cooling (see Appendix A). The formulation of Model A cannot be easily modified if we want to include other mechanisms. We therefore developed a modified version of the model presented by \citetads{2005A&A...443..291S}, using the radiative cooling function derived from AtomDB emissivities with the abundances derived in Section \ref{sec:tshock} and a magnetic dipole geometry as described by \citetads{2005A&A...440..185C}. We explored a range of specific accretion rates, $a$, between 0.2 and 2.0 g s$^{-1}$ cm$^{-2}$, resulting in a range of shock temperatures from 15.9 to 21.3 keV. The models use $R_{\rm in}=2.7~\rm R_{\rm WD}$ \citep{2011MNRAS.411.1317R,2014arXiv1405.0855S}. Tables \ref{tab:2} and \ref{tab:1} list the resulting fluxes and line ratios from this model (Model B), respectively, with a shock temperature of 19.7 keV, which matches the shock temperature derived from the WD mass of 0.79 \MSUNnospace, a specific accretion rate of 0.6 g s$^{-1}$ cm$^{-2}$, and an accretion rate $\dot{M}$=1.74$\times$10$^{-11}$~\rm $M_{\odot}$ yr$^{-1}$. The line ratios are higher than the measurements, especially for low-Z elements. Line fluxes from H-like 
ions are greater than the measurements, while He-like lines are in closer agreement with the data. Although a model with a lower accretion rate would help to match the H-like line fluxes, there is a trade-off between matching these and the He-like lines fluxes and the continuum, which tends to be underestimated by models with very low $\dot{M}$.  

\begin{table*}
\caption[]{H/He line ratios, measured and derived from the models discussed in Section \ref{sec:model}.} 
\label{tab:1}
\centering
\begin{tabular}{lcccccccc}
 \hline\hline
& O& Ne& Mg& Si& S& Ar& Ca& Fe\\
\hline
\hbox to 1in{Data                    \leaders\hbox to 0.4em{\hss.\hss}\hfill}& 2.95$\pm$0.25& 2.01$\pm$0.10& 1.67$\pm$0.08& 1.26$\pm$0.04& 1.23$\pm$0.11& 0.74$\pm$0.14 & 1.02$\pm$0.31& 0.37$\pm$0.06\\
\hbox to 1in{Model A\tablefootmark{a}\leaders\hbox to 0.4em{\hss.\hss}\hfill}&       4.22&       2.64&       2.71&      2.19&       1.68&       1.16&       1.05&       0.50 \\ 
\hbox to 1in{Model B\tablefootmark{b}\leaders\hbox to 0.4em{\hss.\hss}\hfill}&       4.96&       2.95&       2.91&      2.43&       1.93&       1.26&       1.17&       0.56 \\
\hbox to 1in{Model C\tablefootmark{c}\leaders\hbox to 0.4em{\hss.\hss}\hfill}&       2.30&       1.67&       1.65&      1.47&       1.27&       1.00&       0.94&       0.48 \\
\hbox to 1in{Model D\tablefootmark{d}\leaders\hbox to 0.4em{\hss.\hss}\hfill}& 3.07$_{-1.00}^{+0.09}$& 2.22$_{-0.14}^{+0.26}$& 1.98$_{-0.15}^{+0.07}$& 2.11$_{-0.01}^{+0.10}$& 1.58$_{-0.21}^{+0.13}$& 1.17$_{-0.09}^{+0.04}$&1.00$_{-0.16}^{+0.04}$& 0.53$_{+0.08}^{-0.30}$ \\
\hline
\end{tabular}
\tablefoot{Same as Table \ref{tab:2}}
\end{table*}

\subsection{Varying the elemental abundances in the cooling function}
\label{sec:abund}

In the analysis above, bremsstrahlung dominates the cooling between $\log{T_{\rm shock}}$ and about $\log{T} = 7.3$, since it gives the observed EMD and the relative H-like ion intensities (Table \ref{tab:2}). Given that the low-Z He-like line fluxes are underpredicted, one possibility is that the cooling rate below $\log{T}=7.3$ is lower than the adopted cooling rate.  One way of reducing the cooling rate below $\log{T} = 7.3$ is to reduce the elemental abundances, thereby reducing the emission-line cooling from K-shell Mg, Si, and S and L-shell Fe, which dominate the cooling function for $\log{T} < 7.3$ \citepads{2012ApJ...756..128F}. A model with reduced abundances can indeed reproduce the observed line ratios if the elemental abundances are about $\sim$10\% of the solar abundances (see Model C in Tables \ref{tab:2} and \ref{tab:1} and Fig. \ref{fig:ratios}).  However, since the bremsstrahlung continuum is not sensitive to these abundances, the line-to-continuum ratios of the H-like and He-like lines are drastically reduced and the observed spectrum is far from  adequately fit.

\subsection{Cooling model with thermal conduction}
\label{sec:conduction}

Another possibility is that extra heat is added in the regions of the column that are closer to the WD surface, thereby increasing the fluxes of the low-Z He-like lines.  We discuss several possible mechanisms in Appendix A, along with several mechanisms that could potentially alter the line ratio diagnostics.  We conclude that photoionization, steady-flow departures from pressure equilibrium, resonant scattering, Compton cooling, and electron-ion temperature differences are unlikely to explain the discrepancy. A dipole geometry and the gravitational potential difference between the top and bottom of the accretion column also fail to explain the discrepancy, as shown by the models discussed above.  An additional possibility is thermally unstable cooling \citepads{1984ApJ...280..313I,1992A&A...256..660I}.  Thermal conduction seems promising in that it transfers heat from the region above $10^7$ K, where the H-like lines are formed, to the cooler regions where the He-like lines of O, Ne, Mg, and Si are produced.  

In clusters of galaxies, thermal energy from outer regions can be transported to the central cooling gas by conduction, but its importance in this context has been considered doubtful by several authors \citepads{1981ApJ...247..464B, 1983ApJ...267..547T, 1986ApJ...306L...1B}.  Early studies assumed that the magnetic fields strongly suppress conduction perpendicular to the field. Later, \citetads{2001ApJ...562L.129N} showed that a turbulent magnetohydrodynamic medium could support a conductive heat flux, which would play a significant role in the energy equation.  In nonmagnetic accreting WDs, \citetads{2008A&A...483..231L} proposed that the X-ray spectra of dwarf nova can be fit with a model where the material in the innermost portion of the accretion disk is evaporated into a flow, forming a corona that surrounds the WD.  They found that only small conductivities, of about 1\% of the Spitzer value, are able to provide acceptable fits, presumably because the strong shear in the boundary layer produces magnetic fields nearly perpendicular to the temperature gradient.
 
In magnetic CVs, cooling flow models generally ignore thermal conduction because its effect is modest in the high-temperature part of the flow, reducing the emission measure by a few percent  above a few times $10^7$ K \citepads{1984ApJ...280..313I}.  However, that energy is comparable to the thermal energy of the gas below $10^7$ K.  The combination of lower temperature and higher radiative cooling coefficient creates steep temperature gradients in that region, and the divergence of thermal conduction is more important.  If a large fraction of the conduction flux were deposited between about $2 \times 10^6$ and $10^7$ K, it would alleviate the discrepancy in the H-like to He-like ratios of low-Z elements.

We computed simple models of a constant pressure flow cooling from a shock temperature of 19 keV with a constant mass flux.  That is, $n$ is inversely proportional to $T$, and $v$ is inversely proportional to $n$ (eq. \ref{eq17}). The choice of density is not important because the cooling rate is proportional to the density squared, while the cooling length scale is inversely proportional to density, and the divergence of the conductive flux is proportional to the inverse of the distance scale squared.  We describe the previous discussion in the following equations:

\begin{equation}
P_{\rm rad} \propto n_{\rm e}^{2}\ ,
\end{equation}

\begin{equation}
dx \propto n_{\rm e}^{-1}\ ,
\end{equation}

\begin{equation}
\nabla F_{c} \propto  dx^{-2} \propto n_{\rm e}^{2}\ ,
\end{equation} 

\noindent
where $F_{c}$ is the standard Spitzer conductive flux and $P_{\rm rad}$ is the radiative cooling rate. Thus, we have a one-parameter family of models provided that the shock jump conditions determine the relation between flow speed and maximum temperature. 

We solved the equation for thermal energy  by specifying a temperature grid and determining the distance needed to cool from one temperature to the next.  If thermal conduction is ignored, we obtain a simple linear equation (see Section \ref{sec:cf}).  With thermal conduction, the equation is cubic.  

If $\Delta t_i$ is the time needed to cool $\Delta T_i = T_i -T_{i+1}$ and $V_i$ is the flow speed, then the thickness of that cell is

\begin{equation}
 \Delta x_i = V_i \Delta t_i
,\end{equation}

\noindent
and the contribution of that cell to the emission is $n_i^2 \Delta x_i$. We find

\begin{equation}
\Delta t_i = (5 n_i k_B \Delta T_i \Delta x_i + \Delta Fc_i)/(n_i^2 \Lambda \Delta x_i)
,\end{equation}

\noindent
where $\Delta Fc_i$ is the difference in thermal conduction fluxes entering and leaving cell $i$. That yields the equation

\begin{equation}
n_i^2 \Lambda \Delta x_i^3 - 5n_iVk_B \Delta T_i \Delta x_i^2 - V Fc_i \Delta x_i + V \kappa T_i^{2.5} \Delta T_i = 0
,\end{equation}

\noindent
where $\kappa$ is the thermal conduction coefficient.

The effect of conduction is to transfer a few percent of the energy from the hot part of the flow to the cooler regions, increasing the emission measure at lower temperatures.    Nearly all the conductive flux is deposited below $10^6$ K, however, and the EMD at higher temperatures is almost indistinguishable from the model without conduction.

\subsection{The empirical EMD}
\label{sec:poa}

We can empirically derive  an upper limit on the EMD  using the observed emission line fluxes, their tabulated emissivities, and a fitting algorithm. To derive the empirical EMD, we use a Markov-Chain Monte Carlo process as implemented in {\it PINTofALE\/} \citepads[PoA;][]{2000BASI...28..475K}. A detailed explanation of the method, its potential, and its limitations can be found in \citetads{1998ASPC..154..844K}. Given the nature of the method, the shape of the EMD depends on the emission lines chosen for fitting. Basically, the EMD is obtained by fitting the minimum of the loci curves. The loci curves are constructed by measuring the fluxes of emission lines using the relationship

\begin{equation}
 EM(T)=\frac{4\pi d^2 f_{\lambda}}{A \epsilon_{\lambda}(T)}\ ,
\label{eq:loci}
\end{equation}

\noindent
where $d$ is the distance to the source, $f_{\lambda}$ is the flux of the emission line in ergs s$^{-1}$ cm$^{-2}$, $A$ is the element abundance, and $\epsilon_{\lambda}(T)$ is the line emissivity as a function of temperature in units of ergs cm$^3$ s$^{-1}$. Although the derived EMD depends on the assumed elemental abundances, PoA has the option of reconstructing an abundance-independent EMD using the measured line ratios of H- and He-like lines of O, Ne, Mg, Si, S, Ar, Ca, and Fe. The resulting best-fit solution for the EMD is shown in Fig. \ref{fig:emd}, while the range of fluxes and line ratios are listed in Tables \ref{tab:2} and \ref{tab:1} as Model~D.  The overall trend of the empirical EMD agrees with the physically-based EMDs derived above. The empirical EMD mostly differs from  theoretical EMDs in  temperatures where no emission lines were measured (e.g., $\log{T} \gtrsim$ 8.1 or $\log{T} \lesssim$ 6.6) and there are only a few continuum, line-free regions (e.g., 2.2-3.2\AA).

\section{Discussion and conclusions}
\label{sec:disc}

We used the high-quality spectra obtained with {\it Chandra\/} HETG from the intermediate polar EX~Hya to test models for the cooling of the postshock flow, often applied to the X-ray emission from the accretion column of magnetic CVs. We have found that the fluxes of the H-like lines are well reproduced by models of an isobaric cooling flow, while fluxes of low Z He-like lines are underestimated, suggesting that either the cooling in the low-temperature region of the accretion column does not remove heat efficiently, or there is an additional source of heat. On the other hand, the X-ray data from clusters of galaxies show less low-temperature plasma than expected from the similar isobaric cooling flow scenario.  Thus in galaxy clusters there must be an  extra heat source that is strong enough to prevent the plasma from cooling below 10$^{7}$ K \citepads[e.g.,][]{2014Natur.515...85Z}.

We studied different variables that could affect the shape of the EMD and the emission line fluxes, such as a dipole geometry, gravity, and chemical abundances. Neither a dipole geometry nor the gravitational potential difference between the top and bottom of the column are able to modify the EMD in such a way as to match the observed line fluxes and continuum simultaneously, and because most of the continuum is due to the bremsstrahlung emission at the shock temperature, it is possible that the accretion column is not tall enough for these effects to be significant. A subsolar set of elemental abundances allows us to reproduce the observed line ratios, but the line fluxes are greatly underestimated as are the line-to-continuum ratios. Thermal conduction, which has been considered in clusters of galaxies, does not resolve the discrepancy.

\begin{figure}
\includegraphics[width=\columnwidth]{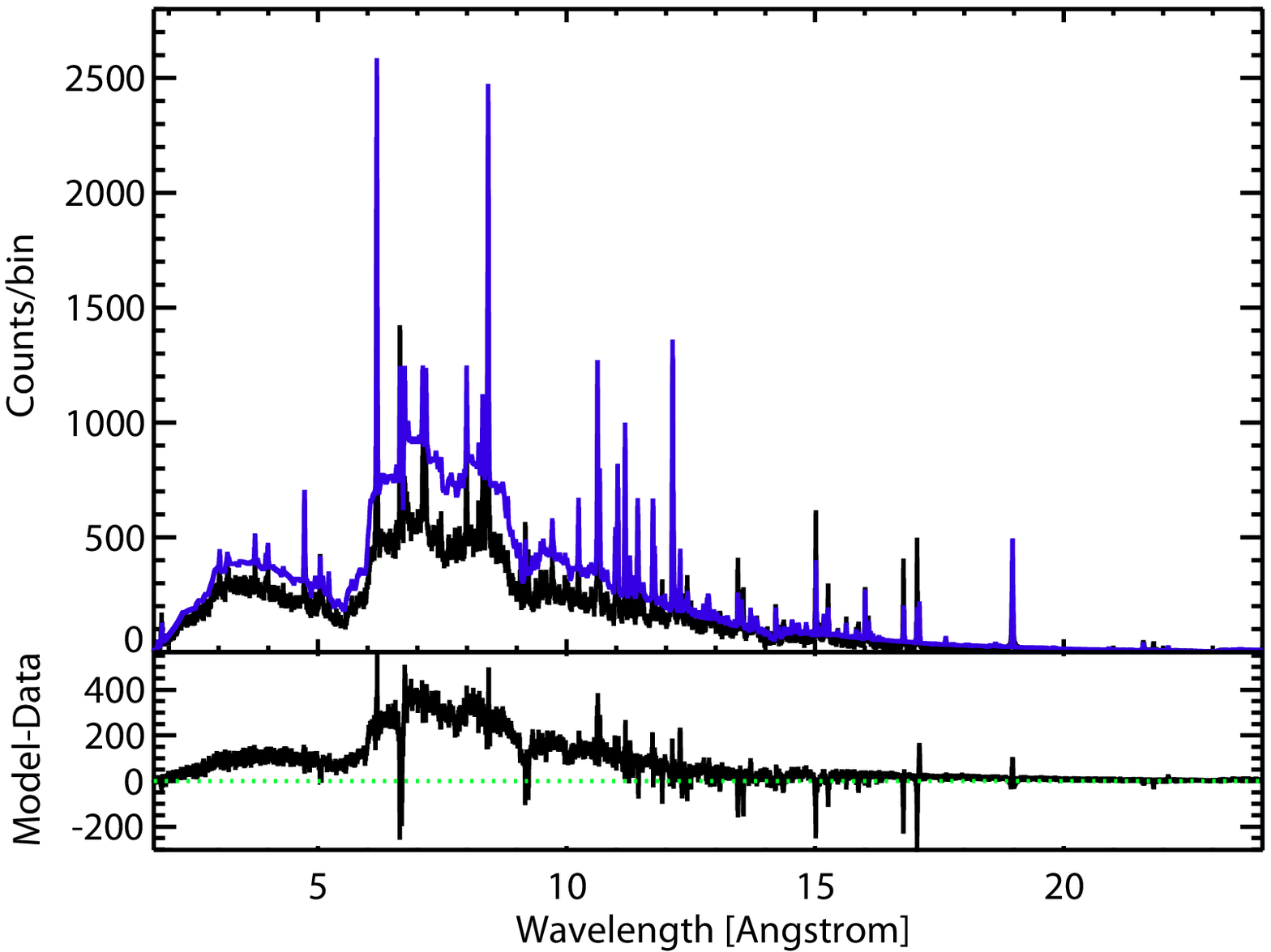}
\includegraphics[width=\columnwidth]{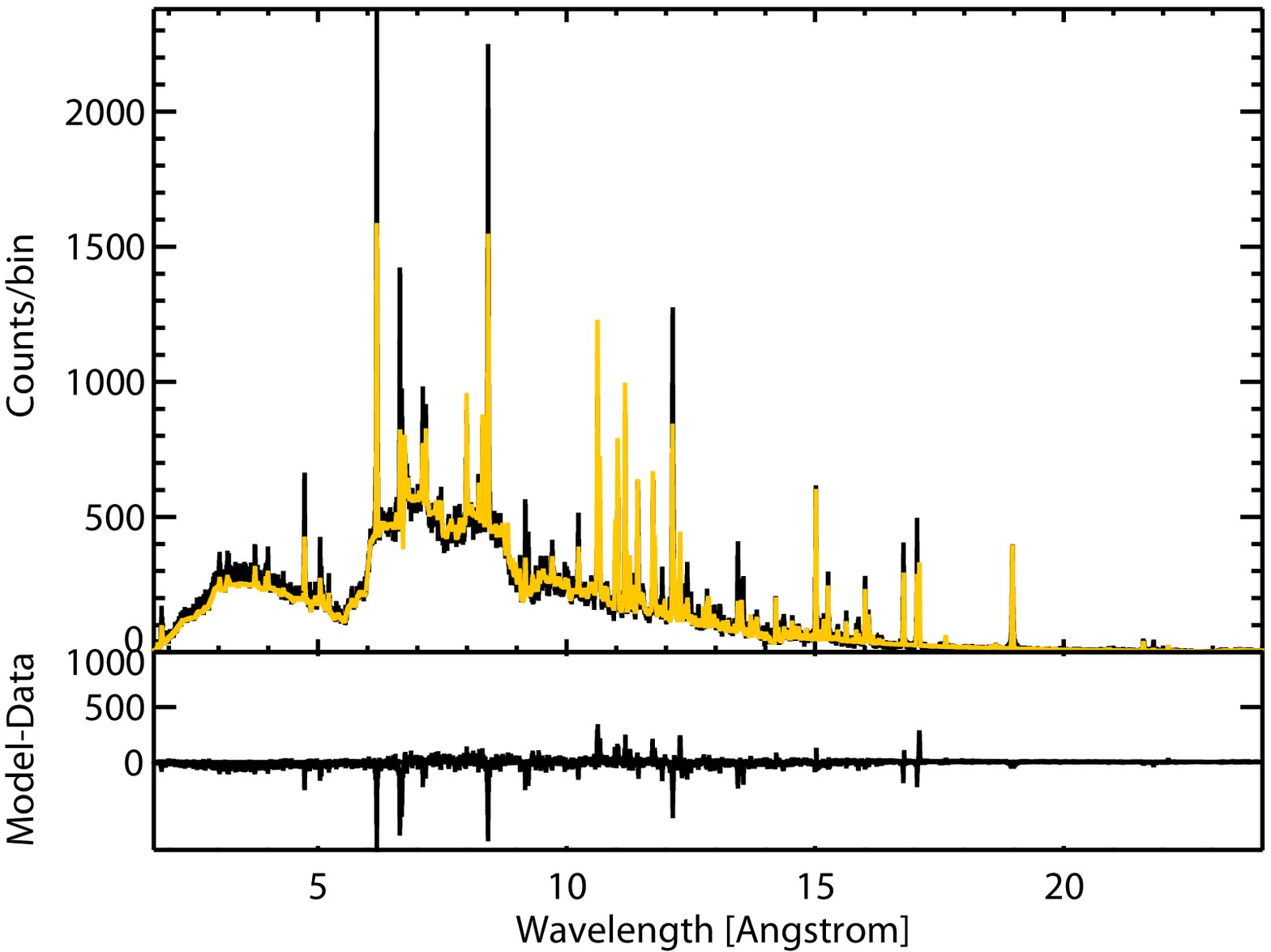}
 \caption{Comparison between the {\em Chandra} MEG$\pm$1 spectrum (black full line) and spectra obtained from the different models discussed in Section \ref{sec:model}. {\em Top:} Accretion column with a magnetic dipole geometry and change of gravitational potential with height (blue full line, Model B, see Section \ref{sec:dipole} for details). {\em Bottom:} EMD derived from the fit of the loci curves obtained from the flux measurements (orange full line, Model D, see Section \ref{sec:poa} for details). The {\em \textup{model-data}} panels show that the continuum emission from model B overestimates the data. Given that all the spectral models were derived assuming the same accretion rate, $\dot{M}$=1.74$\times$10$^{15}$ g s$^{-1}$, model B requires lower accretion rates to match the data. } 
 \label{fig:spec_poa}
 \end{figure}

A comparison of the predicted spectra from the different models discussed in Section~\ref{sec:model} and the {\em Chandra} observation is shown in Fig.~\ref{fig:spec_poa}. The spectra from Models A and C are not shown in this figure because Model A reproduces the results already shown in Figure \ref{fig:spinmax}, while Model C greatly overestimates the continuum and underestimates the line fluxes.  Spectrum from Model B computed for an accretion rate of $\dot{M}$=1.74$\times$10$^{-11}$~\rm $M_{\odot}$ yr$^{-1}$ overestimates the continuum (Fig. \ref{fig:spec_poa}, upper panel), indicating that the accretion rate from this model should be lower that the accretion rate found in the simple isobaric cooling flow model (model A). Model D (Fig. \ref{fig:spec_poa} lower panel) makes a reasonable prediction of the observed spectrum and line ratios, although it underpredicts the fluxes of emission lines in the $\approx$5-10 \AA~\rm region.

The appendix discusses several other possible mechanisms that could modify the EMD and shows that they are unlikely to account for the discrepancies between the model and the observation for the flux of the He-like lines.  Thermal conduction does not alleviate the problem.  A possible explanation is the expected thermal instability of the cooling flow \citepads{1984ApJ...280..313I,1982ApJ...258..289L}, so that the shock height and shock speed vary with time.  Hydrodynamical simulations will be needed to determine whether such instabilities can change the H-like to He-like intensity ratios by the observed amount, but they should change the ratios in the correct sense.  First, the nonlinear thermal instability causes the hot part of the flow to cool at constant density rather than constant pressure.  That means that the energy each particle must lose is 3/2 kT rather than 5/2 kT.  Second, the energy that is not converted to radiation in the constant density cooling produces secondary shock waves that reheat the gas.  The secondary shocks should have speeds of a modest fraction of the intial shock speed, and they could plausibly produce temperatures of a few million K, leading to extra emission in He-like lines of the lower-Z elements.  This simple picture is probably complicated by random 
variations in the density of the gas striking the shock.

Model D shows that the empirical EMD can be rather complex and better theoretical models will be needed  to understand the high-resolution X-ray spectra of magnetic accreting WDs that will be obtained with ground-breaking future instruments such as the calorimeter onboard ASTRO-H.

\begin{acknowledgements}

We thanks Vinay Kashyap, Randall K. Smith, and Adam Foster.  G.J.M.L. acknowledge support from: NASA to the Smithsonian Astrophysical Observatory (SAO) under Chandra GO7- 8026X; grants PICT 2011/269 (Agencia) and PIP D-4598/2012 (CONICET/Argentina). C.W.M.'s contribution to this work was performed under the auspices of the U.S. Department of Energy by Lawrence Livermore National Laboratory under Contract DE-AC52-07NA27344. V.S. was supported by German Research Foundation (DFG) grant WE 1312/48-1.

\end{acknowledgements}
\bibliographystyle{aa}    
\bibliography{listaref}

\appendix
\section{Modifications to the cooling model}

There are several physical processes that must operate at some level in the shock-heated plasma and that would in principle affect the relative intensities of the H-like emission lines.

{\it \textup{\textup{\textit{Photoionization:}}}}  X-rays from the hotter parts of the flow  heat and ionize the cooler parts. X-ray emission from the photoionized accretion column above the shock was identified by \citetads{2010ApJ...711.1333L} in several lines in the form of very large velocity width components.  In the shocked gas, the importance of photoionization is greatly reduced because of much higher density.  We have estimated the effect of photoionization on the cooling flow by computing models of steady-flow shocks with the model of \citetads{1979ApJS...39....1R}.  The atomic rates in this model have been updated over the years, but they are generally not as accurate as the AtomDB rates \citepads[see][for a comparison]{2001ApJ...556L..91S}.  However, the percentage changes due to including processes such as photoionization should be reliable.  Simple 3000 $\rm km~s^{-1}$ shock wave models with and without photoionization showed no difference in the H-like to He-like line ratios.

{\it Departure from constant pressure:}  The shocked plasma slows from 1/4 the shock speed to zero as it settles to the WD surface.  That implies a pressure gradient that can influence the emission measure distribution.  However, models with magnetic support cool at nearly constant density down to $10^6$ K, and therefore nearly constant velocity. These models show changes in the line ratios at only a percent level compared to models without magnetic support.  
The perpendicular  component of the magnetic field alone can  support the cooling plasma and  the field is entirely parallel to the flow in the standard shock model.  However, observations of supernova remnants show strong amplification of the magnetic field, quite likely by an instability that produces perpendicular field \citepads[e.g.,][]{2003ApJ...584..758V,2004MNRAS.353..550B}, and this kind of mechanism may operate in accretion shocks as well.

{\it Thermal instability:} Larger changes in the EMD could result from the thermally unstable nature of the postshock cooling region \citepads{1983ApJ...268..291I,1982ApJ...258..289L}.  The instability causes oscillations in the shock height and shock speed and the cooling gas can develop large low-pressure regions and subshocks \citepads[e.g.,][]{1992A&A...256..660I}.  It is not obvious whether thermal instability would increase or decrease the He-like to H-like ratio, or how big the effect would be.  Hydrodynamic models are needed.

{\it Resonant scattering:}  Terada et al. (2001, 2004) investigated the effects of scattering in the resonance lines in magnetic CVs.  They
found that resonance scattering produced asymmetric emission and modulation on the spin period in polars, but not in intermediate
polars.  \citetads{1997ApJ...474..774F} also showed that resonant scattering was not important in EX~Hya.

{\it Electron-ion temperature difference:}  If the shock wave heats ions, but not electrons, as is observed to occur in collisionless shocks in supernova remnants \citepads{2013SSRv..178..633G}, the electrons are gradually heated by Coulomb collisions.  The result is that the peak electron temperature is a bit lower than the nominal shock temperature, and the electrons  stay near that peak temperature longer than in a single temperature flow.  Based on the shock models mentioned above, this increases the Fe XXVI line flux relative to the others, but not very strongly.

{\it Compton Cooling:}  Compton cooling can compete with bremsstrahlung if the density is low, reducing the bremsstrahlung emission at the highest temperatures.  For a WD temperature of 25{,}000 K \citepads{2002A&A...382..984E} and the densities indicated by the Fe XVII and Fe XXII lines \citepads{2001ApJ...560..992M,2003ApJ...588L.101M}, however, the Compton cooling rate is less than 1\% of the bremsstrahlung cooling rate. Compton scattering is an 
additional cooling mechanism, which can potentially decrease the plasma temperature in the upper hot layers  and, therefore, decrease the fluxes of  H-like lines. However, this cooling mechanism is effective only for massive WDs ($> 1$ \MSUNnospace) and high local mass accretion rates ($a > 1~\rm g~\rm s^{-1}~\rm cm^{-2}$) \citepads{2008A&A...491..525S}.


\end{document}